\documentclass[acmsmall]{acmart}
\usepackage{subfig}
\usepackage{multirow}
\usepackage{color}
\definecolor{ForestGreen}{RGB}{34,139,34}

\AtBeginDocument{%
  \providecommand\BibTeX{{%
    \normalfont B\kern-0.5em{\scshape i\kern-0.25em b}\kern-0.8em\TeX}}}

\setcopyright{acmcopyright}
\copyrightyear{2018}
\acmYear{2018}
\acmDOI{10.1145/1122445.1122456}

\acmPrice{15.00}
\acmISBN{978-1-4503-XXXX-X/18/06}

\begin{document}

\title{Multi-Cycle-Consistent Adversarial Networks for Edge Denoising of Computed Tomography Images}

\author{Xiaowe Xu}
\authornote{Both authors contributed equally to this research.}
\author{Jiawei Zhang}
\affiliation{%
  \institution{Guangdong Cardiovascular Institute, Guangdong Provincial Key Laboratory of South China Structural Heart Disease, Guangdong Provincial People's Hospital, Guangdong Academy of Medical Sciences}
  \streetaddress{106 Zhongshan Second Road}
  \city{Guangzhou}
  \state{Guangdong}
  \postcode{510080}
}
\email{xuxiaowei@gdph.org.cn}
\orcid{0000-0002-1046-6379}
\author{Jinglan Liu}
\authornotemark[1]
\author{Yukun Ding}
\author{Tianchen Wang}
\email{jliu16@nd.edu}
\affiliation{%
  \institution{Department of Computer Science and Engneering, University of Notre Dame}
  \city{Notre Dame}
  \state{IN}
  \postcode{46556}
}

\author{Hailong Qiu}\authornotemark[1]
\author{Haiyun Yuan}
\author{Jian Zhuang}
\author{Wen Xie}
\authornote{Corresponding authors.}
\affiliation{%
  \institution{Department of Cardiovascular Surgery, Guangdong Cardiovascular Institute, Guangdong Provincial Key Laboratory of South China Structural Heart Disease, Guangdong Provincial People's Hospital, Guangdong Academy of Medical Sciences}
  \streetaddress{106 Zhongshan Second Road}
  \city{Guangzhou}
  \state{Guangdong}
  \postcode{510080}
}
\email{zhuangjian5413@163.com}
\author{Yuhao Dong}
\author{Qianjun Jia}
\authornotemark[2]
\author{Meiping Huang}
\authornotemark[2]

\affiliation{%
  \institution{Department of Catheterization Lab, Guangdong Cardiovascular Institute, Guangdong Provincial Key Laboratory of South China Structural Heart Disease, Guangdong Provincial People's Hospital, Guangdong Academy of Medical Sciences}
  \streetaddress{106 Zhongshan Second Road}
  \city{Guangzhou}
  \state{Guangdong}
  \postcode{510080}
}
\email{huangmeiping@126.com}

\author{Yiyu Shi}
\email{yshi4@nd.edu}
\affiliation{%
  \institution{Department of Computer Science and Engneering, University of Notre Dame}
  \city{Notre Dame}
  \state{IN}
  \postcode{46556}
}

\renewcommand{\shortauthors}{X. Xu and J. Liu, et al.}

\begin{abstract}
As one of the most commonly ordered imaging tests,
computed tomography (CT) scan comes with inevitable radiation
exposure that increases the cancer risk to patients.
However, CT image quality is directly related to radiation dose,
thus it is desirable to obtain high quality CT images with as little dose as possible. 
CT image denoising tries to obtain
high dose like high-quality CT images (domain $Y$) from low
dose low-quality CT images (domain $X$), 
which can be treated as an image-to-image
translation task where the goal is to learn 
the transform between a source domain $X$ (noisy images) and 
a target domain $Y$ (clean images). Recently, 
cycle-consistent adversarial denoising network (CCADN) 
has achieved state-of-the-art results by enforcing 
cycle-consistent loss without the need 
of paired training data, since the paired data is 
hard to collect due to patients' interests and 
the cardiac motion. 
On the other hand, 
out of concerns on patients' privacy and data security, protocols typically require clinics to perform medical image processing tasks including CT image denoising locally, i.e., edge denoising.
Therefore, the network models needs to achieve high performance under various computation resource constraints including memory and performance. 
Our detailed analysis of CCADN raises a number of 
interesting questions which point to potential ways to further improve its performance
 using same or even less computation resources. 
For example, if the noise is large leading 
to significant 
difference between domain $X$ and domain $Y$, 
can we bridge $X$ and $Y$ with a intermediate 
domain $Z$ such that both the denoising process 
between $X$ and $Z$ and that between $Z$ and $Y$ are 
easier to learn? As such intermediate domains 
lead to multiple cycles, how do we best enforce 
cycle-consistency? Driven by these questions, 
we propose a multi-cycle-consistent adversarial 
network (MCCAN) that builds intermediate domains and 
enforces both local and global cycle-consistency for edge denoising of CT images. 
The global cycle-consistency couples all generators together to model the whole denoising process, 
while the local cycle-consistency imposes effective 
supervision on the process between adjacent domains. 
Experiments show that both local and global 
cycle-consistency are important for the 
success of MCCAN, which outperforms CCADN in terms of denoising quality with slightly less computation resource consumption.
\end{abstract}



\maketitle

\section{Introduction}
The privacy and security of patient data have always been the primary concern in medical applications among hospitals and clinics. 
As such, protocols typically require medical image processing tasks such as denoising \cite{shan20183,wolterink2017generative,chen2017low,yang2018low}, segmentation \cite{xu2019whole,wang2020ica, liu2019machine,xu2018quantization,mdunet,conzhang2021,pyranet}, and diagnosis \cite{xu2020imagechd} to be performed locally, i.e., on the edge.
However, local machines and devices are usually with rather limited computation resources including memory capacity and performance compared with those in the cloud. The constrained resources can have 
profound impact on the design of medical image processing algorithms. 
In this paper, we will use Computed tomography (CT) image denoising as a vehicle to demonstrate it. 

CT is one of the most widely used 
medical imaging modality for showing anatomical structures \cite{you2018structurally}. The foremost 
concern of CT examination is the associated exposure 
to radiation, which is known to increase the lifetime 
risk for death of cancer \cite{hobbs2018physician}. 
The radiation dose can be lowered at the cost 
of increased noise
\cite{li2014adaptive,you2018structurally}. 
Such noise in CT image leads to both degraded perceptual
quality and degraded diagnostic confidence of a doctor. A
general principle in dose management in practice is “as low
as reasonably achievable” \cite{mayo2014managing}. Thus the resulted 
images are denoised for minimized the loss on perceptual quality 
and diagnostic confidence of radiologists. 
Even with tremendous effort 
and significant progresses in the past few decades, the
radiation exposure of CT scan was still estimated to account
for up to two percent of cancer in United States 
\cite{sodickson2009recurrent}.


Various deep neural network (DNN) based methods
exist for CT image 
denoising \cite{shan20183,wolterink2017generative,chen2017low,yang2018low}, which require paired clean 
and noisy images for training. 
Yet paired images are hard to collect 
due to patients' interests and 
the cardiac motion. Therefore, 
simulations are usually used to generate 
such paired data, where the simulated noise patterns
can be different from the real ones, leading to 
biased training results \cite{kang2018cycle}. 
To address this issue, recently 
cycle-consistent adversarial denoising 
network (CCADN) was 
proposed in \cite{kang2018cycle}, which 
formulates CT image denoising as an 
image-to-image translation 
problem without paired training
data.
CCADN consists of two generators: 
One transforms noisy CT images (domain $X$) 
to clean ones (domain $Y$) and 
the other transforms clean CT images (domain $Y$) 
to noisy ones (domain $X$). 
Both generators are trained by adversarial 
loss. In addition, they form a cycle where a noisy 
CT image can 
be transformed to a clean one and transformed back 
to a noisy one (i.e., $X\rightarrow Y\rightarrow X$).
Cycle-consistency loss is 
defined by the difference between the two
noisy CT images, which is a key 
component to control the training 
of both generators for better performance. 
Cycle-consistency loss is also imposed for 
$Y\rightarrow X\rightarrow Y$ transform. 
However, CCADN only works well when the noise levels are low. This is 
due to the fact that it only contains two domains $X$ and $Y$ and therefore 
its efficacy degrades as the noise becomes stronger, leading to larger
differences between $X$ and $Y$ that are harder to learn. A larger neural network 
with stronger representation power is needed, which may not be feasible 
with the limited computation resources on the edge. 

\begin{figure}[!tb]
\setlength{\belowcaptionskip}{-10pt}
\centering
\includegraphics[width=0.65\textwidth]{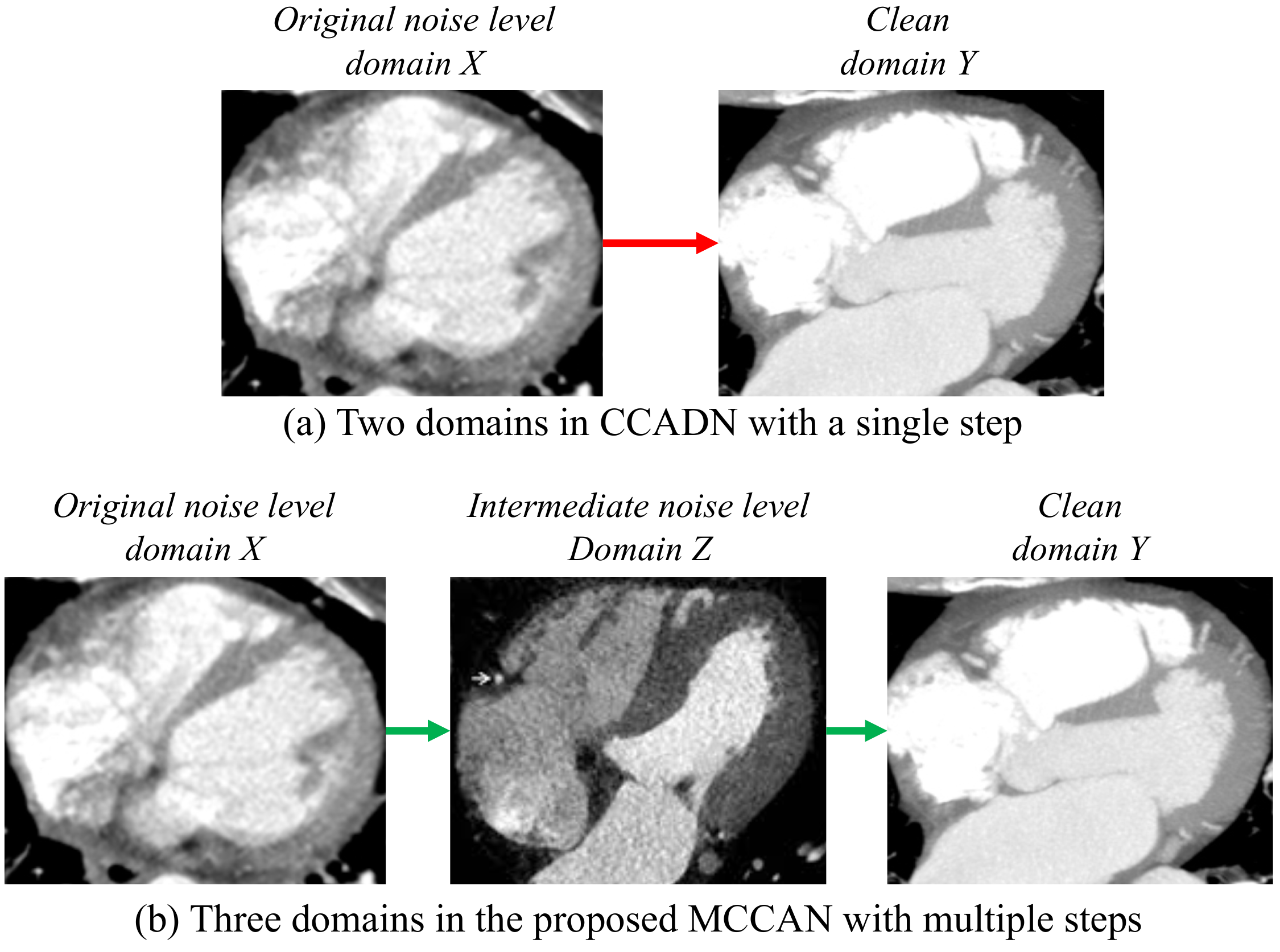}
\caption{{\color{black} Comparison of domians in (a) CCADN and (b) the proposed MCCAN. CCADN performs single-cycle-consistent adversarial training with two domains, while the proposed MCCAN performs multiple-cycle-consistent adversarial training with more than two domains, e.g., three. }}
\label{fig:example}
\end{figure}

To enhance the performance of CCADN without increasing the resource consumption, 
 as shown in Fig. \ref{fig:example}, we propose to
establish an intermediate domain between the original 
noisy image domain $X$ and clean image domain 
$Y$, 
and decompose the denoising task into multiple 
coupled steps such that each step is easier 
to learn by DNN-based models.
Specifically, we construct an additional domain $Z$ 
with images of intermediate noise level between 
$X$ and $Y$. These images can be 
considered as a step stone in the 
denoising process and provide additional 
information for the training of the denoising network. 
The multi-step framework particularly suits the 
denoising problem: while it is difficult to 
either find or define a good collection of 
images in the ``half-cat, half dog" domain 
in ``cat-to-dog'' type of image translation 
problems, a domain $Z$ of images with intermediate
level of noise exist naturally.
\textcolor{black}{
In addition, when the denoising problem (two domains) is divided into several subproblems ($N$ domains), each subproblem is much easier to solve. 
In this way, the network complexity for each subproblem (e.g., number of parameters) is usually lower than $1/(N-1)$ of the original network \cite{xu2018scaling}.
Thus, the overall computation consumption can be reduced.
}

With the new domain $Z$, we further propose a 
multi-cycle-consistent adversarial network to 
perform the multi-step denoising, which 
builds multiple cycles of different 
scales (global cycles and local cycles) between the 
domains while 
enforcing the corresponding cycle-consistencies. 
Specifically, global cycles combine all the 
generators and domains together to model the entire 
denoising process. The local cycles serve
two purposes.
First, they impose effective 
supervision on the generators between adjacent domains. 
Second, while each step is easier, the multi-step 
framework leads to deeper networks and makes it 
challenging for end-to-end training. The local 
cycles can provide gradient from the supervised 
training of easier tasks on shallower networks 
and thus alleviating the problem.
The experimental results show that both 
global cycles and local cycles are 
necessary, and our method MCCAN outperforms the state-of-the-art competitor CCADN with a slightly less resource consumption.

\section{Related Works}
\subsection{CT Image Denoising} Numerous CT image denoising 
methods can be categorized as three types: signogram filtering-based method, 
iterative reconstruction, and image space denoising \cite{you2018structurally}. 
The first two types of methods are usually embedded within 
the CT scanner as commercial algorithms, thus we focus more 
on the last type of methods for research. 

Signogram filtering-based method perform in the original 
projection space before filtered backprojection is applied 
to reconstruct images \cite{manduca2009projection,liu2017discriminative}. 
One common advantage of these method is the noise properties in 
projection space are fairly well-understood. However, the image 
sharpness may degrade because the edges are not well-defined in 
projection data \cite{li2014adaptive}.

Iterative reconstruction are considered the most accurate one 
by using statistical assumptions about the noise properties in 
projection space, prior information in image space, and various 
accurate information of the specific scanner \cite{nuyts1998iterative}. 
However, the implementation highly depend on specific scanner models 
and is very computational extensive for each scan \cite{xu2007real,li2014adaptive}.

Image space denoising is performed on the reconstructed images and 
thus the computation cost is much lower than that in the first two 
category. In recently years, deep neural networks and various method 
developed in other area are combined with CT image denoising including GAN, 
autoencoder, perceptual loss, transfer learning, 
3D convolution \cite{shan20183,wolterink2017generative,chen2017low,yang2018low}.
Mostly recently, CycleGAN is applied to CT denoising as CCADN and achieves 
better results than state-of-the-art \cite{kang2018cycle}.

It can be hard to find a standard metric to measure the denoising 
performance when there is no paired samples for test. For the protection 
of patients and operators, repetitive CT scan is usually not permitted due 
to the additional radiation dose. 
Even if repetitive scan is available, the cardiac motion or the changed operating 
condition will make two scans different. 
This problem is alleviated by simulating corresponding 
low dose images from high dose images with noise modeling \cite{karmazyn2009ct,green2018learning}. 
However, noise should be added in the sinogram domain in the synthetic 
CT scan images, which is too difficult to implement without the 
assistance from the CT scanners' vendor \cite{kang2018cycle}. 
Besides, the additional noise pattern can be different from the 
real noise pattern. This will introduce bias in the data and end 
up with biased denoising models. 
\subsection{Image-to-Image Translation}
Our work are closely related to some of the popular 
image-to-image translation models using generative 
adversarial networks \cite{isola2017image} or neural 
style transfer \cite{johnson2016perceptual}. 
Image-to-image translation also includes some other 
artifact removal problems similar to denoising such as 
raindrop removal and shadow removal \cite{qian2018attentive,liu2018erase,wang2018stacked}.
\cite{wang2018stacked} uses a joint-learned two-step 
approach for shadow removal where one conditional 
GAN \cite{mirza2014conditional} is used to detect the 
shadow region and the result is used by another 
conditional GAN for shadow removal. However these two steps 
are mostly specific to a small set of problems and 
can not be applied to other cases with more steps.

While the using of cycle consistency loss has achieved 
significant progresses \cite{zhu2017unpaired,yi2017dualgan,kim2017learning}, 
these models still have some drawbacks. CycleGAN often succeeds 
on translation of low level features including color and texture 
but has little success on tasks with geometric changes \cite{zhu2017unpaired}. 
We anticipate that the multiple-step approach can potentially 
alleviate this problem. On the other hand, CycleGAN can be 
inefficient for translation in multiple domains, because the 
number of translation model grows quadratically with the number 
of domains. Motivated by this, StartGAN and 
ComboGAN \cite{choi2017stargan,anoosheh2018combogan} propose 
new models to get better scalability. The training method of 
CycleGAN in the facial attribute transfer experiment 
in \cite{choi2017stargan} is essentially the MCCAN without global 
cycles. The ComboGAN is used to perform multi-step 
transformation (e.g. changing gender after changing hair color). 
The results show that the image quality degrade with more 
translation steps. We analyze the possible reasons, the lack of 
global cycles, in Section~\ref{method}. 


Coupled GAN (CoGAN) is proposed in \cite{liu2016coupled} to learn 
a joint distribution in different domains without paired samples. 
Based on the assumption that deep neural networks learn a 
hierarchical feature representation, CoGAN enforces the GANs 
to decode high-level semantics in the same way by sharing the 
weights. In order to translate a image $x$ in domain $X$ to 
domain $Y$, it have to find the random vector that generates 
$x$ through the generator for $X$ and they apply the generator 
for $Y$ to this random vector. Such a search process could be 
very time consuming \cite{gatys2016image}. Another limitation 
is the transformation only success when $x$ is covered by the 
generator for $X$ (can be generated by this generator). While 
it is not discussed in their paper, we found that their approach 
implicitly uses a feature map domain as a bridge for the translation. 
Specifically, if we assume the query image is always covered 
by the corresponding generator, their structure can be consider 
as building a domain graph what each images domain is connected 
to a central feature map domain. When doing the translation, 
it always translate as $X\to Z\to Y$ where $Z$ is the feature 
map domain. 
In this paper, we equip MCCAN with available CT images 
from different domains (radiation dose) directly for the denoising 
task while without increasing the resource consumption. 

\section{Multi-Cycle-Consistent Adversarial Networks }
\label{method}

Given training images that are either labelled as noisy 
(domain $X$) or clean (domain $Y$), we first construct 
a new domain $Z$ which contains images with an 
intermediate noise level between $X$ 
and $Y$. How to obtain $Z$ is flexible in practice. 
It can either be obtained from $X$ and $Y$ by 
separating out those images with intermediate noise
level if available, or from images scanned 
with medium dose, 
or from common techniques 
including injecting intermediate level of noise 
to the images in $Y$. 

\begin{figure}[htb]
\subfloat[]{
\begin{minipage}[b]{0.5\linewidth}
\label{fig:compa}
\includegraphics[height=1in]{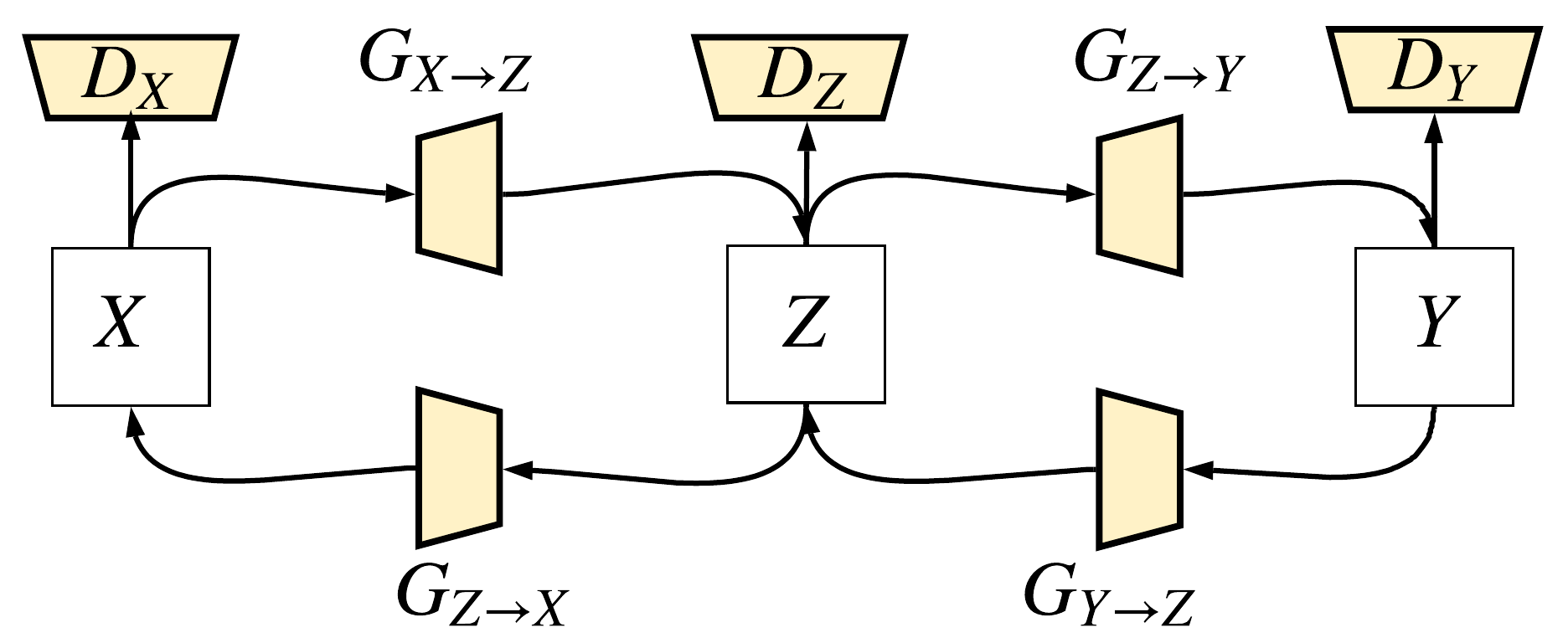}
\end{minipage}
}
\subfloat[]{
\begin{minipage}[b]{0.5\linewidth}
\label{fig:compb}
\includegraphics[height=1in]{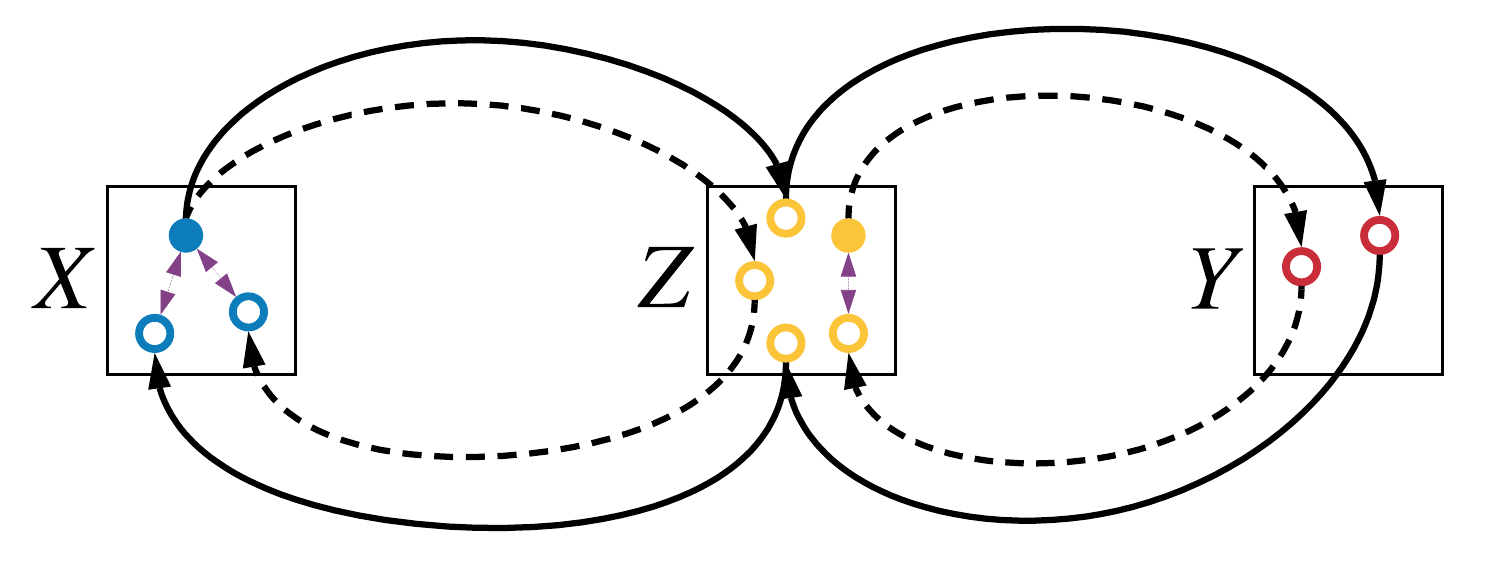}
\end{minipage}
}
\caption{ (a) Structure of MCCAN and (b) its cycles. The arrows inside each domain denote the computation of cycle-consistency loss. The solid and dashed arrows across domains form global and local cycles, respectively. For clarity, we only show cycles from left to right. Symmetric cycles going from right to left also exist but are not shown.}
\label{fig:comp}
\end{figure}

With CT images from three domains, 
the multi-step 
denoising architecture of MCCAN 
is shown in Fig. \ref{fig:compa}. We train four 
convolutional neural networks as generators and 
three as discriminators. 
Arrows in Fig. \ref{fig:compa} define how images 
are transformed in the training stage. Specifically, 
the generator $G_{X\to Z}$ aims 
to transform an image from $X$ to $Z$. 
$G_{Z\to X}$, 
$G_{Z\to Y}$, and $G_{Y\to Z}$ can be 
interpreted similarly. Discriminators 
$D_X$, $D_Y$, 
and $D_Z$ aim to distinguish the ``real'' 
images originally belonging to the domains 
$X$, $Y$, and $Z$ respectively 
from the ``fake'' images transformed from 
other domains.

As the MCCAN structure in 
Fig. \ref{fig:compa} contains 
thee domains, there are multiple 
ways in which we can construct 
cycles (paths where an image from a 
source domain is transformed through 
one (in \cite{zhu2017unpaired}) 
or several other domains (in this paper) 
and back to the source domain) for 
cycle-consistent loss. 
In particular, we introduce two types of 
cycles as shown in Fig. \ref{fig:compb}. 
In this figure, each dot represents an 
image, which is color-coded based on the domain. 
The solid ones represent the images 
originally in the domain 
(``real'' ones), and the hollow ones 
represent those transformed 
from another domain (``fake'' ones). 
As such, the dashed arrows form the 
{\bf local cycles}, each of which 
goes across only two adjacent domains. 
On the other hand, the solid arrows 
constitute a {\bf global cycle} that 
starts from $X$ through 
$Z$, $Y$, $Z$, and back to $X$ sequentially. 
Note that in the figure we only show 
half of the cycles (from left to right) 
for clarity, and the other half which 
are from right to left and symmetric 
to the ones shown also exist. 
We then enforce cycle-consistency 
loss, which measures 
the difference between the original 
images and the final images 
produced at the end of the cycle as represented by 
the small 
arrows within each domain in Fig~\ref{fig:compb}. 
Ideally, the images transformed 
back to the source domain should 
be identical to the original images. 
The cycle-consistency loss is 
applied to every cycle, no matter 
whether it is local or a global.

The global 
cycles are important for the denoising 
performance due to the following reason. 
In the inference stage, an input noisy 
CT image $x$ in domain $X$ will be transformed 
by $G_{X\to Z}$ and 
$G_{Z\to Y}$ sequentially, which 
means $G_{X\to Z}$ and 
$G_{Z\to Y}$ are coupled by data 
dependency. Without global cycles, 
$G_{X\to Z}$ and $G_{Z\to Y}$ will be trained 
independently. The global cycles enable the 
joint training of the generators, which models 
the denoising path used in the inference 
stage for better consistency. 

The local 
cycles are also important to address two issues in the training. 
First, the global cycles go through 
all the four generators and 
have long paths for the gradient 
to back-propagate, which makes the 
end-to-end optimization 
difficult. The locals cycles are shallow and have 
shorter paths for the gradient to back-propagate. Second, 
adversarial training only enforces 
the generators to output 
``fake'' images identically distributed as 
the original ``real'' images in the 
intermediate domain $Z$. 
However, they do not necessarily preserve 
the meaningful content in the inputs, 
which is critical for the denoising task. 
The local cycle-consistency supervises each generator 
to learn to transform images while preserving their 
meaningful content from the inputs more easily.

\begin{figure*}[htb]
\centering
\subfloat[]{
\begin{minipage}[b]{0.48\linewidth}
\centering
\label{fig:compexpa}
\includegraphics[height=1in]{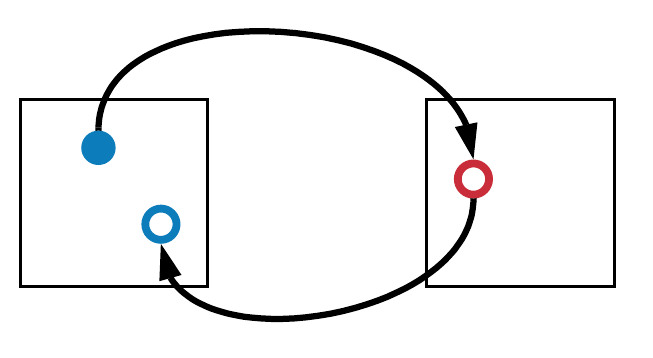}
\end{minipage}
}
\subfloat[]{
\begin{minipage}[b]{0.48\linewidth}
\centering
\label{fig:compexpb}
\includegraphics[height=1in]{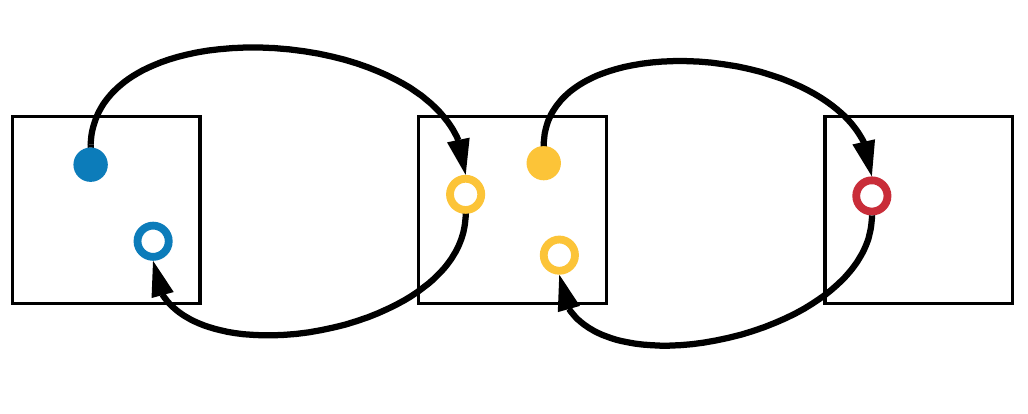}
\end{minipage}
}
\quad
\subfloat[]{
\begin{minipage}[b]{0.48\linewidth}
\label{fig:compexpc}
\centering
\includegraphics[height=1in]{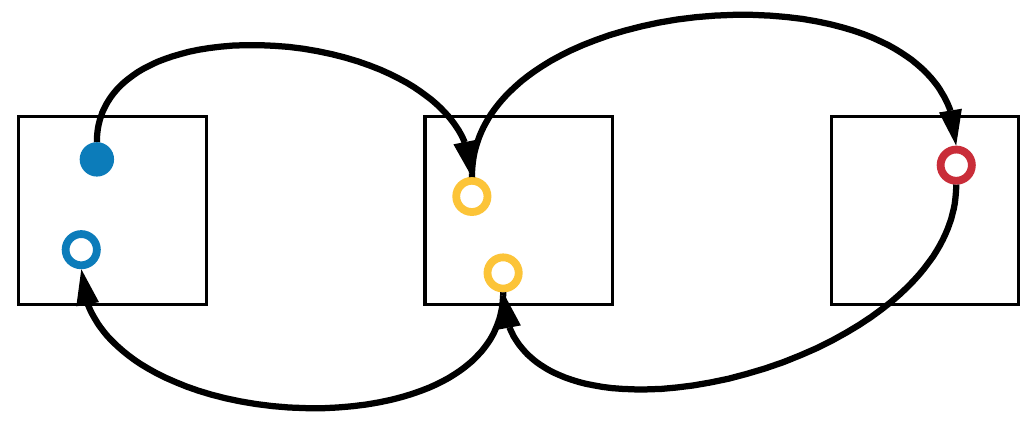}
\end{minipage}
}
\subfloat[]{
\begin{minipage}[b]{0.48\linewidth}
\label{fig:compexpd}
\centering
\includegraphics[height=1in]{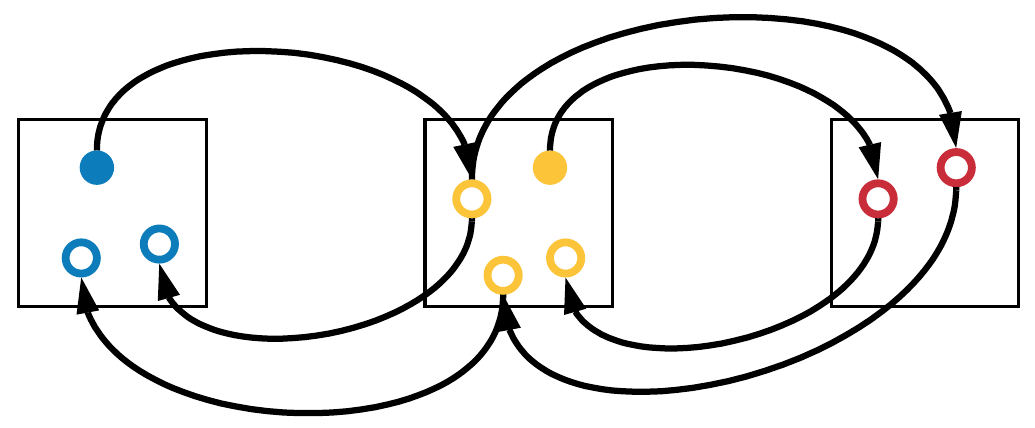}
\end{minipage}
}
\caption{Comparison of (a) CCADN, (b) MCCAN without global cycles (c) MCCAN without local cycles, and (d) MCCAN. For the clarity of presentation, we only show cycles from left to right and symmetric cycles from right to left also exist.}
\label{fig:compexp}
\end{figure*}

In summary, our MCCAN has three major advantages 
over CCADN. First, it decomposes the one-step transform 
into multiple steps using constructed images in a 
intermediate domain as a step stone. Second, 
it not only incorporates global cycles 
that model the denoising path in 
the inference stage for consistency, 
but also uses local 
cycles that provide strong supervision to facilitate the more challenging training process. 
Third, the network structure of generators can be simplified due to the relatively easier task in each transformation, thus potentially reducing memory and computation consumption.

Note that in the discussion so far, only one 
intermediate domain was assumed. It is also 
possible to include more than one intermediate 
domains with more global and local cycles. However, 
our study suggests that any additional domains beyond 
one will not introduce further performance 
gain in the dataset we explored.

\subsection{Network Architectures}
We compare MCCAN with a state-of-the-art 
CT denoising framework 
CCADN \cite{kang2018cycle}. 
In order to see how the local cycles and 
global cycles contribute to the final 
performance, we also implement and compare 
MCCAN without local cycles 
and without global cycles 
respectively as ablation study. 
The various structures are 
shown in Fig. \ref{fig:compexp}. 
or the clarity of presentation, 
only cycles from left to right are 
shown, but symmetric cycles from 
right to left also exist.

\begin{figure*}[htb]
\centering
\subfloat[]{
\begin{minipage}[b]{1\linewidth}
\label{fig:structureb}
\centering
\includegraphics[height=1.13in]{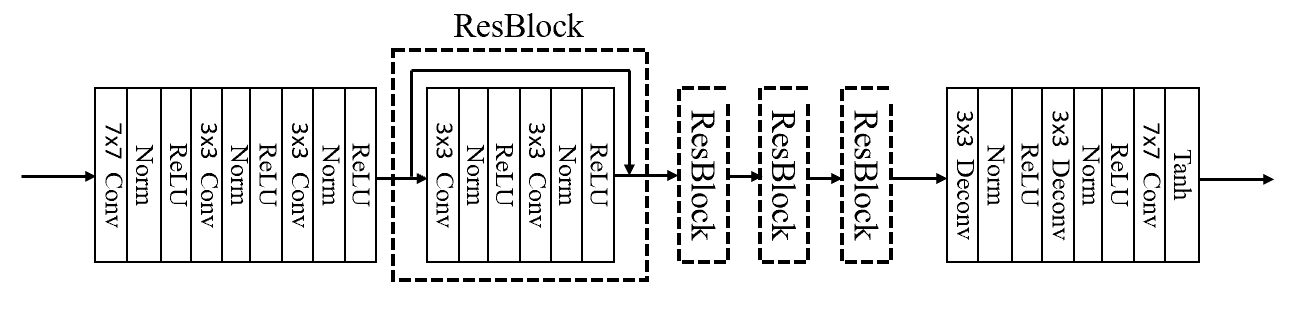}
\end{minipage}
}
\\
\subfloat[]{
\begin{minipage}[b]{1\linewidth}
\label{fig:structurea}
\centering
\includegraphics[height=0.73in]{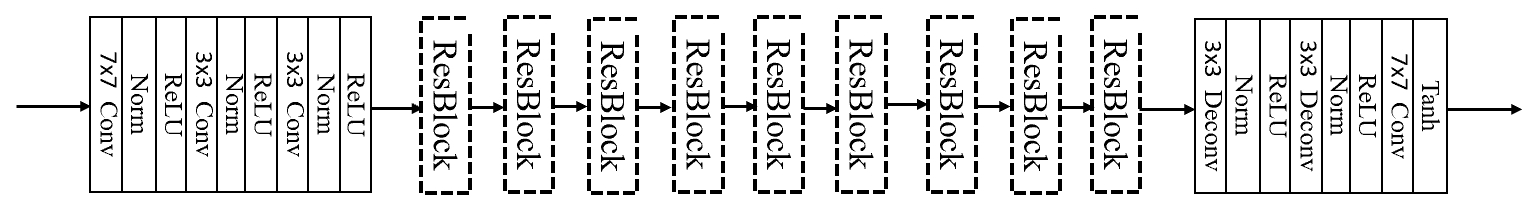}
\end{minipage}
}
\caption{Network architectures of generators for (a) MCCAN, and (b) CCADN. ResBlock represents residual blocks \cite{he2016deep}. For MCCAN, MCCAN without local cycles, and MCCAN without global cycles, the network architecture in (b) are shared. Because every generator in such frameworks, the generator only transfer between adjacent domains. Nevertheless, generators in CCADN need to transfer images between more than one domains, thus with larger networks. }
\label{fig:structures}
\end{figure*}
To ensure that model sizes, computation operations, and number of training epochs are the same 
for fair comparisons, different 
network structures are applied 
in experiments. Generally speaking, 
all discriminators are used for a 
same discrimination task in each 
image domain, so all discriminators 
share a same network structure. 
However, generator structures are 
different for different tasks, 
since the capability of generators 
denoising by one step and the capability 
of generators denoising by more steps, 
two step here, are supposed to 
be different.

Traditional convolution networks are 
used as the discriminator for all domains 
in all experiments. In terms of the generators, 
since there are two types of generators in 
our experiments, different network architectures 
are utilized in our experiments. Both network 
architectures share the same layers for 
pre-processing and post-processing, as shown 
in Fig. \ref{fig:structures}. 
However, the generators transferring images
images between two domains that are more different 
from each other should have more layers and 
parameter, and vice versa. So the generators in 
CCADN is implemented with more residual blocks 
(ResBlock) as shown in 
Fig. \ref{fig:structurea}, 
while generators in MCCAN, MCCAN without local cycles, 
and MCCAN without global cycles are all implemented 
as the network architecture shown in 
Fig. \ref{fig:structureb}. 
As a result, the total number of parameters used 
in each experiment in inference is around 11M to 
make sure that each method consumes the same memory resources.

\subsection{Training Objectives}
Finally, we state the training objective used in our framework. 
Denote $\{G\}$ and $\{D\}$ as the set of generators 
and discriminators respectively. 
Denote $I\in \{X,Y,Z\}$ as one domain and $D_I$ as the 
discriminator associated with domain $I$. 
We let $C_i$ be a cycle and $P_{i,j}$ be a path 
of half $C_i$ that has the same source domain, 
where $i, j$ are used to distinguish different 
cycles and paths merely. 
For example, $X\rightarrow Z\rightarrow X$ is 
a cycle, saying $C_1$, 
thus we can have $P_{1,1} = X \rightarrow Z$, 
and $P_{1,2} = Z \rightarrow X$, which are both 
half cycles of $C_1$. 
$\{P_{I}\}$ represents the 
set of all the paths that end at 
domain $I$. We denote $I_{C_i}$ as the 
source domain of $C_i$ and 
$G_{P_{i,j}}$ as the ordered function composition 
of the generators in $P_{i,j}$. Thus, 
the total adversarial loss is

\begin{equation}
  \label{equ:loss_gan_xy}
  \mathcal{L}_{GAN}(\{G\}, \{D\}) =\sum_{ I\in\{X,Y,Z\}} \sum_{ P_{i,j}\in\{P_I \}}\mathcal{L}_{GAN}(I,P_{i,j})
\end{equation}

where $\mathcal{L}_{GAN}(I,P_{i,j})$ is the adversarial loss associated with domain $I$ and the transform path $P_{i,j}$. $\mathcal{L}_{GAN}(I,P_{i,j})$ is obtained by
\begin{equation}
\label{equ:loss_gan_xy_i}
\begin{aligned}
  \mathcal{L}_{GAN}(I,P_{i,j}) =
  & \mathbb{E}_{y\sim p_{data}(I)}
  [\log{D_{I}(y)}] \\
  & + \mathbb{E}_{x\sim p_{data}(I_{C_i})}
  [log(1-D_{I}(G_{P_{i,j}}(x)))]
\end{aligned}
\end{equation}
where $p_{data}$ is the distribution of 
``real'' images in a domain and $D_I(x)$ 
represents the probability determined by 
$D_I$ that $x$ is a ``real'' image from 
domain $I$ rather than a ``fake'' one 
transformed by generators from another 
domain. 

The cycle-consistency loss is associated 
with each $C_i$, defined as
\begin{equation}
  \mathcal{L}_{\text{cyc}}(\{G\},C_i) = \mathbb{E}_{x\sim p_{\text{data}}(I_{C_i})}[|G_{C_i}(x)-x|_1].
\end{equation}
The identity loss is associated 
with each generator in 
${G}$, defined as
\begin{equation}
\begin{aligned}
  \mathcal{L}_{\text{idt}}(\{G\}) = &\sum_{I\in\{X,Y,Z\}}
  \sum_{J\in\{X,Y,Z\},J\neq I}(\\
  &\mathbb{E}_{x\sim p_{\text{data}}(I)}[|G_{J\rightarrow I}(x) - x|_1]).
\end{aligned}
\end{equation}

The final optimization problem we solve in 
the training stage is:
\begin{equation}
\begin{aligned}
  \{G\}^* = &\arg\min_{\{G\}}\max_{\{D\}}( \mathcal{L}_{GAN}(\{G\}, \{D\})\\
  &+\lambda_{\text{cyc}} \sum_{C_i\in \{C\}} \mathcal{L}_{cyc}(\{G\}, C_i)\\
  &+\lambda_{\text{idt}}\cdot\mathcal{L}_{\text{idt}}(\{G\}).
\end{aligned}
\end{equation}
where $\lambda_{\text{cyc}}$ and $\lambda_{\text{idt}}$ are 
set to 10 and 0.5 respectively in our experiments.

\section{Experiments and Results}
\subsection{Experiments Setup}
\label{DataGeneration}

\textcolor{black}{The original dataset contains 200 clean 
(normal-dose) 3D cardiac CT images 
and 200 noisy (low-dose) ones from 
various patients for training, and separate 
11 images for test. The dataset is captured from 6 patients (3 for normal dose and 3 for low dose). } All examinations 
are performed with a wide detector
256-slice MDCT scanner (Brilliance iCT; 
Philips Healthcare) providing 8cm of 
coverage. Each 2D CT image is of size 
512$\times$512, which is then
randomly cropped into 256$\times$256 
for data augmentation. We extract the 
noise pattern from the noisy CT 
images and add them to the clean 
CT images with a weighting factor $\frac{1}{2}$ 
to generate new CT images with the intermediate 
noise level.

Following existing works \cite{arapakis2014using,wolterink2017generative,yang2018low}, 
we use the mean and standard deviation (SD)
of pixels in homogeneous regions of
interest chosen by our radiologists 
to quantitatively judge the quality 
of CT images.
The mean, which reflects substance information,
should be as close to that in the origin 
image as possible, 
and the standard deviation, which reflects 
noise, should be as low as possible.

\begin{figure*}
\centering
\subfloat[Original noisy CT images]{
 \label{fig:overview11}
 \centering
 \includegraphics[width=3.8in]{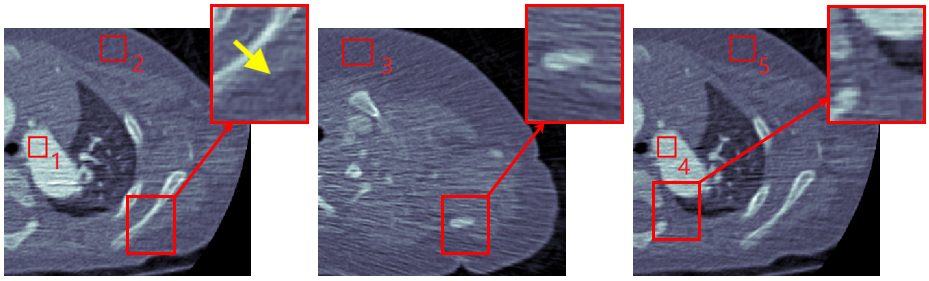}
}
\\
\subfloat[Images denoised by CCADN
]{
 \label{fig:overview12}
 \centering
 \includegraphics[width=3.8in]{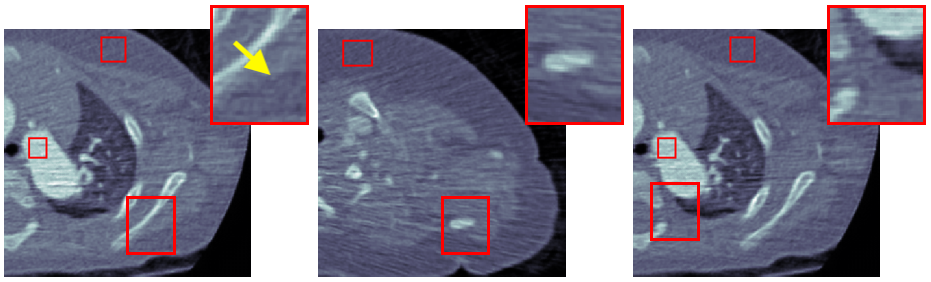}
}
\\
\subfloat[Images denoised by MCCAN without local cycles]{
 \label{fig:overview13}
 \centering
 \includegraphics[width=3.8in]{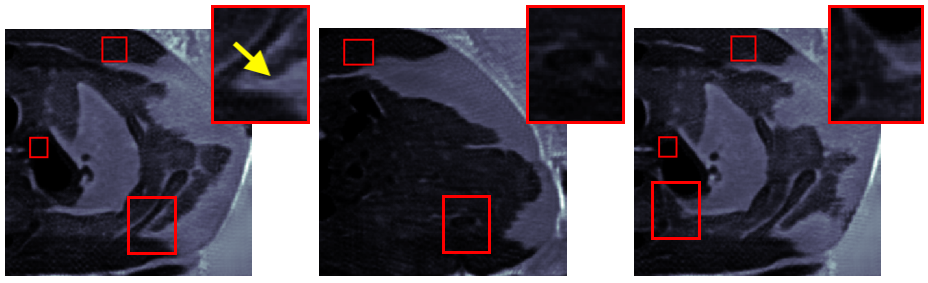}
}
\\
\subfloat[Images denoised by MCCAN without global cycles]{
 \label{fig:overview14}
 \centering
 \includegraphics[width=3.8in]{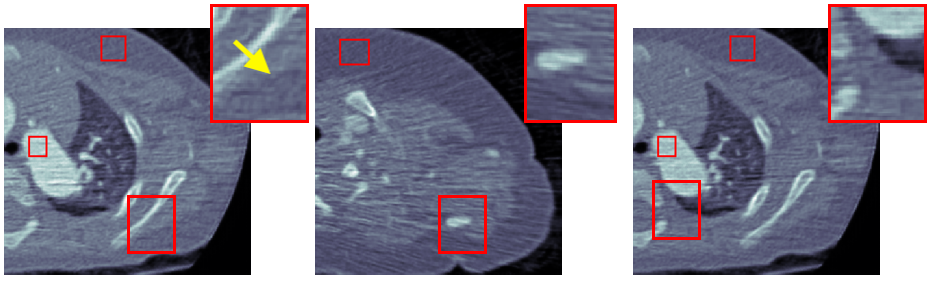}
}
\\
\subfloat[Images denoised by MCCAN]{
 \label{fig:overview15}
 \centering
 \includegraphics[width=3.8in]{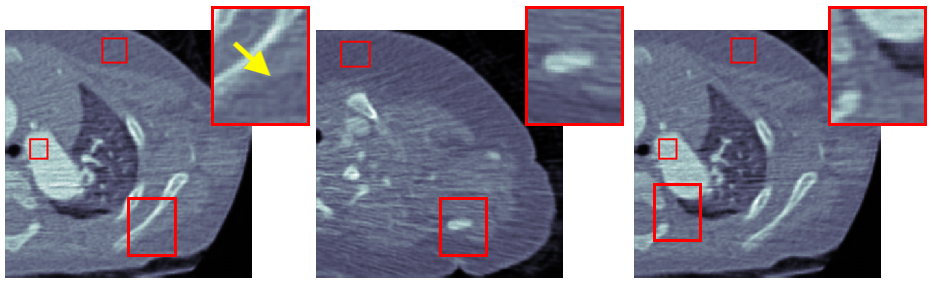}
}
\caption{(a) Original noisy CT images and the corresponding ones denoised by (b) CCADN\cite{kang2018cycle}, (c) MCCAN without local cycles, (d) MCCAN without global cycles, and (e) MCCAN. (Best viewed in color.)}
\label{fig:overview1}
\end{figure*}

\begin{figure*}
\centering
\subfloat[Original noisy CT images]{
 \label{fig:overview21}
 \centering
 \includegraphics[width=3.8in]{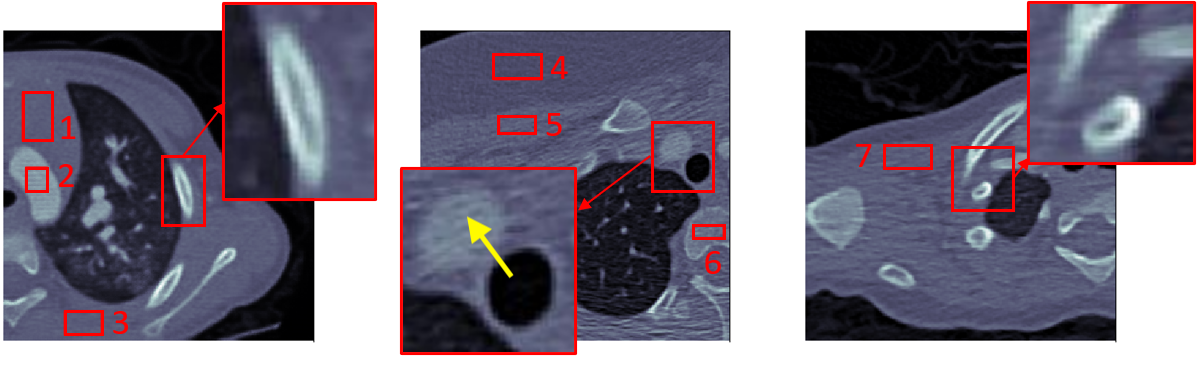}
}
\\
\subfloat[Images denoised by CCADN
]{
 \label{fig:overview22}
 \centering
 \includegraphics[width=3.8in]{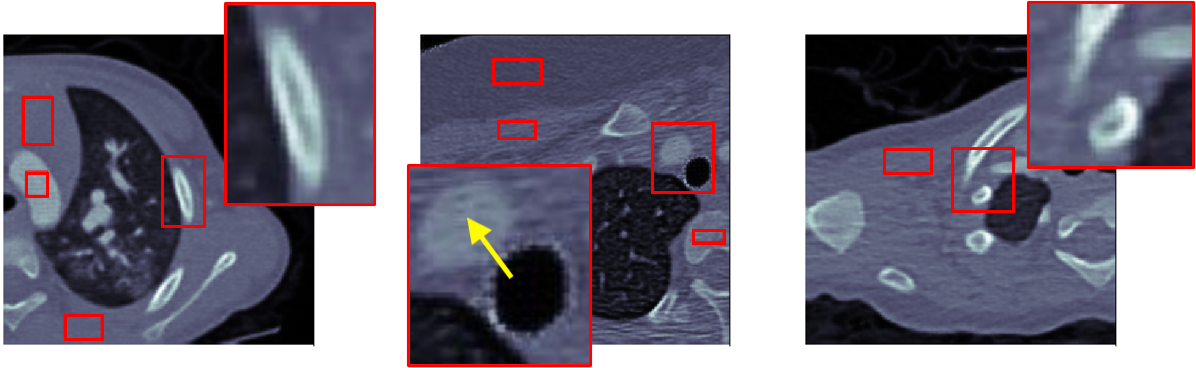}
}
\\
\subfloat[Images denoised by MCCAN without local cycles]{
 \label{fig:overview23}
 \centering
 \includegraphics[width=3.8in]{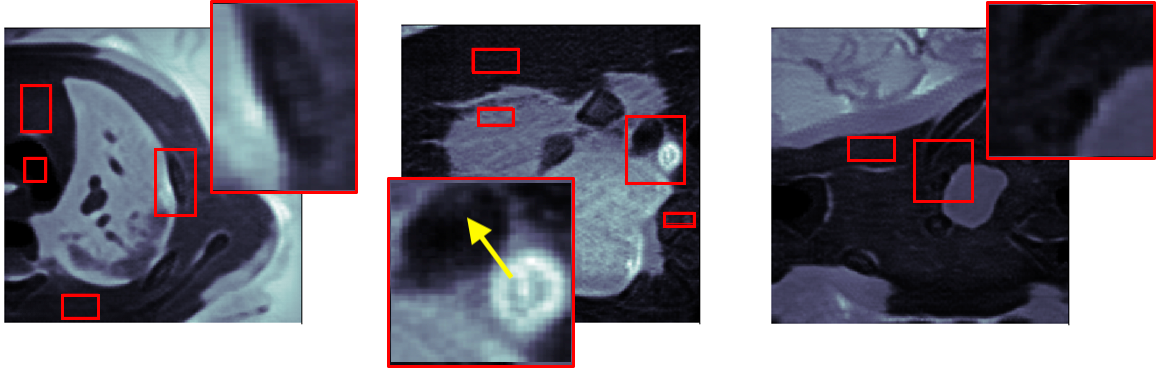}
}
\\
\subfloat[Images denoised by MCCAN without global cycles]{
 \label{fig:overview24}
 \centering
 \includegraphics[width=3.8in]{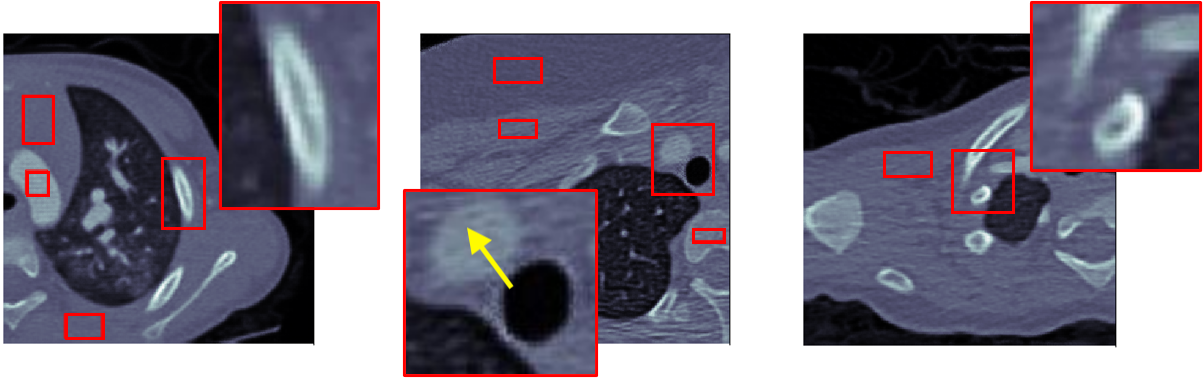}
}
\\
\subfloat[Images denoised by MCCAN]{
 \label{fig:overview25}
 \centering
 \includegraphics[width=3.8in]{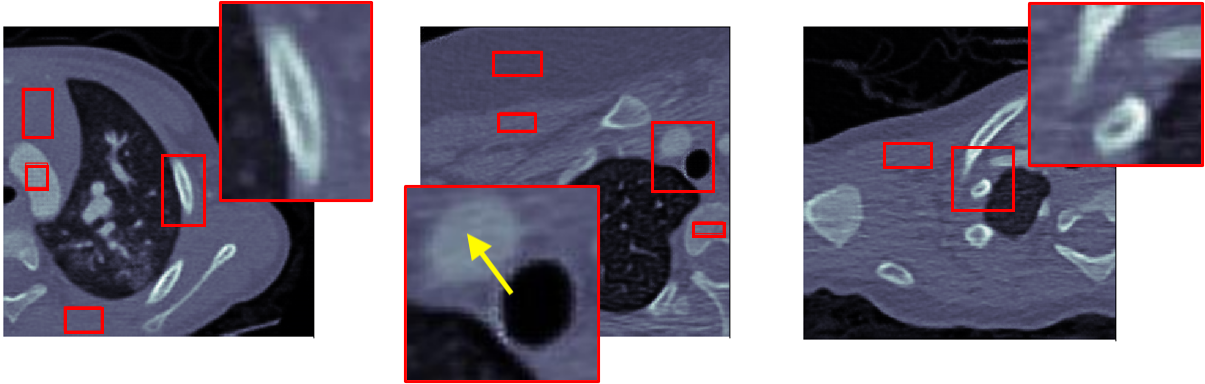}
}
\caption{\textcolor{black}{(a) Original noisy CT images and the corresponding ones denoised by (b) CCADN\cite{kang2018cycle}, (c) MCCAN without local cycles, (d) MCCAN without global cycles, and (e) MCCAN. (Best viewed in color.)}}
\label{fig:overview2}
\end{figure*}

\begin{figure*}
\centering
\includegraphics[width=5.5in]{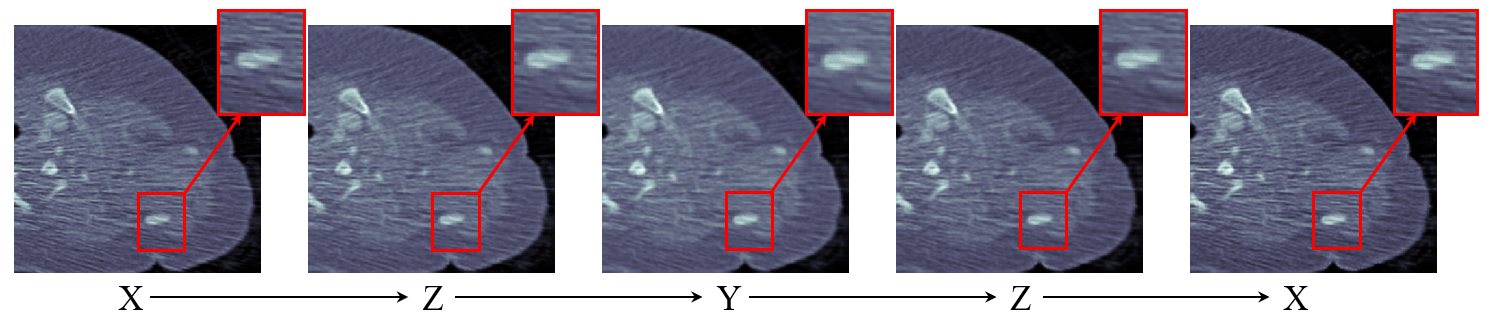}
\caption{An image transformed through X$\rightarrow$Z$\rightarrow$Y$\rightarrow$Z$\rightarrow$X cycle in Fig. \ref{method}. The noise level decreases along X$\rightarrow$Z$\rightarrow$Y and increases along Y$\rightarrow$Z$\rightarrow$X, which conforms to our design.}
\label{fig:ImageInCycle}
\end{figure*}

\begin{table}[tb]
 \begin{center}
 \caption{Comparison of denosing performance between configurations using different number of domains in MCCAN over the selected areas in Fig. \ref{fig:overview11}.}
 \label{table:domains}
 \begin{tabular}{c|c|c}
 \hline
 Method & Mean & SD\\
 \hline
 Original & 1321.2 & 84.5 \\
 \hline
 CCADN (two domains)\cite{zhu2017unpaired} & 1284.1 & 67.8\\
 \hline
 MCCAN (three domains) & 1251.4 & \textbf{60.6}\\
 \hline
 MCCAN (four domains) & 1244.1 & 77.6\\
 \hline
 \end{tabular}
 \end{center}
\end{table}

We first discuss how the number domains affects the denoising performance of MCCAN. Then we compare MCCAN with a state-of-the-art 
CT denoising framework 
CCADN \cite{kang2018cycle}, which is also 
based on cycle-consistency loss but 
contains only two domains. 
In order to see how the local cycles and 
global cycles contribute to the final 
performance, we also implement and compare
MCCAN without local cycles 
and without global cycles 
respectively as ablation study. 
The various structures are 
shown in Fig. \ref{fig:compexp}. 
We train all the networks 
following the setting in 
\cite{zhu2017unpaired}. 
\textcolor{black}{As shown in Fig. \ref{fig:overview1} and Fig. \ref{fig:overview2}, six images chosen by our radiologist are used for the qualitative evaluation, and 12 homogeneous areas annotated by red rectangles and numbered are used for quantitative evaluation.
}
All network sizes and number of 
training epochs are the same 
for fair comparisons.

\subsection{Discussion of Number of Domains}

As shown in Table \ref{table:domains}, the average mean and SD in five areas indicated in Fig. \ref{fig:overview11} are presented.
Compared with the original image, CCADN can largely reduce the noises by about 20\%.
MCCAN with three domains can further improve the image quality with reduced noise by about 8.4\%.
However, MCCAN with four domains obtains reduced improvement compared with that with three domains and CCADN.
Actually this is expected.
More domains require more datasets with different levels of noise, and the difference between the datasets with adjacent levels of noise is smaller. When the difference gets 
too small, the network can no longer learn it effectively, resulting in degraded performance.
For our collected dataset, the optimal number of domains is three, and other datasets may have a different optimal number of domains. How to effectively identify the optimal 
number of domains for a given dataset can be an interesting problem worth further 
studying. 

\subsection{Comparison with the State-of-the-art Method}

\begin{table}[htb]
 \begin{center}
 \caption{Comparison of number of parameters (memory) and operations (computation) between the proposed method and the state-of-the-art CCADN for inference. Note that there is one generator in CCADN, while there are two in MCCAN. FLOP stands for floating point operation.}
 \label{table:paras}
 \begin{tabular}{c|c|c}
 \hline
 Method & Number of parameters & FLOPs\\
 \hline
 CCADN\cite{zhu2017unpaired} & 11.4M & 745G \\
 \hline
 MCCAN w/o local cycles & 11.0M & 668G\\
 \hline
 MCCAN w/o global cycles & 11.0M & 668G\\
 \hline
 MCCAN & 11.0M & 668G\\
 \hline
 \end{tabular}
 \end{center}
\end{table}


\begin{table*}[htb]
 \begin{center}
 \caption{Mean and SD of the selected areas in Fig. \ref{fig:overview1}. }
 \label{table:199_idose_1}
 \begin{tabular}{c|c|c|c|c|c|c|c|c|c|c}
 \hline
 \multirow{2}{*}{Method} & \multicolumn{2}{c|}{Original} & \multicolumn{2}{c|}{CCADN \cite{zhu2017unpaired}} & \multicolumn{2}{c|}{\begin{tabular}[c]{@{}l@{}}MCCAN w/o\\ local cycles\end{tabular}} & \multicolumn{2}{c|}{\begin{tabular}[c]{@{}l@{}}MCCAN w/o\\ global cycles\end{tabular} } & \multicolumn{2}{c}{MCCAN}\\
 \cline{2-11}
 & Mean & SD & Mean & SD & Mean & SD & Mean & SD & Mean & SD\\
 \hline
 Area \#1 & 1942.3 & 118.0
 & 1801.2 & 100.2 
 & 30.9 & 45.2 
 & 1747.7 & 92.6 
 & 1712.8 & {89.8} \\
 \hline
 Area \#2  & 903.3 & 60.0 
 & 928.8 & 47.1 
 & 215.0 & 60.9
 & 932.1 & 46.0 
 & 940.5 & {40.8} \\
 \hline
 Area \#3  & 913.6 & 58.1 
 & 938.2 & 46.0  
 & 96.7 & 41.2 
 & 943.5 & 46.4 
 & 938.0 & {41.3} \\
 \hline
 Area \#4
 & 1944.0 & 132.6 
 & 1821.5 & 103.8 
 & 43.8 & 55.6 
 & 1762.2 & 97.3 
 & 1723.5 & {94.5} \\
 \hline
 Area \#5 & 903.2 & 53.8
 & 930.7 & 42.0
 & 255.0 & 70.7
 & 934.0 & 43.4
 & 942.2 & {36.6}\\
 \hline
 \end{tabular}
 \end{center}
\end{table*}

\begin{table*}[htb]
 \begin{center}
 \caption{\textcolor{black}{Mean and SD of the selected areas in Fig. \ref{fig:overview2}.} }
 \label{table:199_idose_2}
 \begin{tabular}{c|c|c|c|c|c|c|c|c|c|c}
 \hline
 \multirow{2}{*}{Method} & \multicolumn{2}{c|}{Original} & \multicolumn{2}{c|}{CCADN \cite{zhu2017unpaired}} & \multicolumn{2}{c|}{\begin{tabular}[c]{@{}l@{}}MCCAN w/o\\ local cycles\end{tabular}} & \multicolumn{2}{c|}{\begin{tabular}[c]{@{}l@{}}MCCAN w/o\\ global cycles\end{tabular} } & \multicolumn{2}{c}{MCCAN}\\
 \cline{2-11}
 & Mean & SD & Mean & SD & Mean & SD & Mean & SD & Mean & SD\\
 \hline
 Area \#1 & 1067.2 & 24.8
 & 1057.75 & 20.4 
 & 87.2 & 12.1
 & 1058.5 & 20.58 
 & 1061.9 & 19.53 \\
 \hline
 Area \#2   & 1568.4 & 62.4   
 & 1498.4 & 60.6 
 & 47.9 & 22.6 
 & 1504.0 & 59.1 
 & 1547.9 & 54.8 \\
 \hline
 Area \#3  & 1052.9 & 34.4  
 & 1044.1 & 27.9 
 & 56.4 & 12.3 
 & 1044.8 & 28.0 
 & 1047.0 & 24.2 \\
 \hline
 Area \#4
 & 906.6 & 43.2
 & 942.6 & 37.7
 & 120.1 & 25.7  
 & 930.7 & 33.4 
 & 897.1 & 30.9 \\
 \hline
 Area \#5 & 1082.2 & 66.11
 & 1191.3 & 51.8
 & 254.3 & 43.0  
 & 1165.5 & 50.7
 & 1147.3 & 48.8\\
 \hline
 Area \#6 & 1233.0 & 76.9
 & 1398.2 & 65.4
 & 176.0 & 56.3 
 & 1364.6 & 64.4
 & 1367.7 & 63.6\\ 
 \hline
 Area \#7 & 992.5 & 230.4
 & 968.9 & 190.0
 & 340.2 & 67.2 
 & 1010.3 & 188.8
 & 995.8 & 178.3\\
 \hline
 \end{tabular}
 \end{center}
\end{table*}

\subsubsection{Comparison of Resource Consumption}
The comparison of number of parameters and floating point operations (FLOPs) between the proposed method and the state-of-the-art CCADN in inference is shown in
Table~\ref{table:paras}. 
We can notice that the number of parameters and operations of MMCAN is lightly less than that of CCADN, and MCCANs with different configurations have the same number of parameters and operations.
This is because there are two generators with the same network structure in inference for MCCANs with different configurations.
Note that, although MCCAN without 
global cycles is supposed to have 
the same network architecture as 
MCCAN, there is one extra 
discriminator associated with the intermediate domain $Z$. 
The reason for the extra discriminator 
is that if there is only one discriminator 
associated with domain $Z$, the information 
from both domain $X$ and domain $Y$ will 
help it to learn, which results in a 
stronger discriminator than a discriminator 
with only one local cycle. That is, 
information from global scope, both domain 
$X$ and domain $Y$ here, converges at 
the discriminator. Then the information 
would be propagated to generators connected 
with the discriminator as well as whole 
networks. This is what we would not want 
to see. Thus, two discriminators are 
utilized, and one for each local cycle 
to break the global information 
communication. 

\subsubsection{Qualitative Evaluation}
\textcolor{black}{We choose six representative low-dose 
CT images in the test dataset as shown in 
Fig. \ref{fig:overview1} and Fig. \ref{fig:overview2} for qualitative evaluation.
The corresponding denoised images by 
CCADN, MCCAN without local cycles, 
MCCAN without global cycles, and MCCAN are shown in Fig. \ref{fig:overview12}-\ref{fig:overview15} and \ref{fig:overview22}-\ref{fig:overview25}
respectively.}
From the figures we can see that
CCADN can successfully reduce noise in the 
original images. MCCAN without local 
cycles completely fails to produce 
reasonable results. 
A close examination of the images reveal that interestingly the background and the substances are approximately swapped compared with the original images. This is because the high-level features of content distribution are still kept even with such swap, and the discriminator cannot identify the generated image as ``fake'' because of the structure diversity in the training dataset. 
This aligns with 
our discussion on the importance of 
local cycles in Section 2.  
\textcolor{black}{We can take a more closer comparison on the area indicated by yellow arrows in Fig. \ref{fig:overview1} and Fig. \ref{fig:overview2}.
We can observe that compared with the original images and CCADN, only MCCAN can successfully remove the tiny spot indicated by a yellow arrow in Fig. \ref{fig:overview1}.
In addition, MCCAN and MCCAN without global cycles can remove the small hole in Fig. \ref{fig:overview2} which should not exist in the vessel.
}
On the other hand, MCCAN without global cycles 
can successfully 
denoise the image and achieves similar 
quality compared with CCADN. 
This is expected as MCCAN without global 
cycles is essentially formed 
by two cascaded CCADNs. 
\textcolor{black}{Finally, though MCCAN without local or cycles and the complete MCCAN have competitive visual performance with each other, 
the complete MCCAN 
has a relatively smaller noise (less spots, and more smooth boundary) visually.}

To further illustrate the efficacy of the 
MCCAN structure, Fig. \ref{fig:ImageInCycle} 
shows how an image is transformed along a 
global cycle (the path
X$\rightarrow$Z$\rightarrow$Y$\rightarrow$Z$\rightarrow$X). From the figure we can 
see that $X\rightarrow Z \rightarrow Y$ is 
an effective two-step denoising process 
while $Y \rightarrow Z \rightarrow X$ 
incrementally adds noise back. 

\subsubsection{Quantitative Evaluation}

\textcolor{black}{The quantitative results are shown in 
Table \ref{table:199_idose_1} and Table \ref{table:199_idose_2}. 
CCADN can reduce the standard deviation in the 12 
areas by 15\%, 21\%, 
21\%, 22\%, 22\%, 
18\%, 3\%, 19\%, 13\%, 22\%, 15\%, and 18\%, 
respectively}, with resulting 
mean values close to those of the original images. 
Although MCCAN without local cycles 
achieves smallest standard 
deviation in Areas 1, 3 and 4, it leads to 
large mean deviation 
from the original images, which corresponds 
to the structure loss 
in Fig. \ref{fig:overview13}.
\textcolor{black}{MCCAN without global cycles has similar performance 
compared with CCADN, 
with mean values close to original and standard
deviation reduction by 22\%, 23\%, 20\%, 
27\%, 19\%, 17\%, 5\%, 19\%, 23\%, 23\%, 16\%, and 18\%,  respectively. }
Finally, the complete MCCAN behaves the 
best among all 
the methods. 
\textcolor{black}{With mean values close to original, 
the standard deviations are 
decreased by 24\%, 32\%, 29\%, 29\%, 32\%, 21\%, 12\%, 30\%, 28\%, 26\%, 17\%, and 23\%, 
from the original CT images, respectively. }

\section{Conclusions}

In this paper, we propose multi-cycle-consistent
adversarial network (MCCAN) 
for {\color{black}edge denoising of CT images}. MCCAN builds 
intermediate domains and enforces both 
local and global cycle-consistency. 
The global cycle-consistency 
couples all generators together to model the 
whole denoising process, while
the local cycle-consistency imposes 
effective supervision on the denoising process 
between adjacent domains. 
Experiments show that both local and global 
cycle-consistency are important for the 
success of MCCAN, and it outperforms 
the state-of-the-art competitor with slightly less resource consumption. Our code is publicly available.
Considering the practical usage, the computation complexity and the denoising performance still need further improvement, and our future work will focus on optimize the denoising performance while reducing the computation operations at the same time.
\textcolor{black}{In the future work, we will try to apply MCCAN to other medical images such as magnetic resonance imaging (MRI) and ultra sound images, and explore the cycle design theoretically. }


\begin{acks}
This work was supported by the National key Research and Development Program of China (No. 2018YFC1002600), the Science and Technology Planning Project of Guangdong Province, China (No. 2017B090904034, No. 2017B030314109, No. 2018B090944002, No. 2019B020230003), Guangdong Peak Project (No. DFJH201802), the National Natural Science Foundation of China (No. 62006050).

\end{acks}

\bibliographystyle{ACM-Reference-Format}
\bibliography{sample-base}










\end{document}